\documentclass[sigconf]{acmart} 

\usepackage{algorithmic}
\usepackage{graphicx}
\usepackage{adjustbox}
\usepackage{textcomp}
\usepackage{xcolor}
\def\BibTeX{{\rm B\kern-.05em{\sc i\kern-.025em b}\kern-.08em
    T\kern-.1667em\lower.7ex\hbox{E}\kern-.125emX}}
\usepackage[per-mode = fraction]{siunitx}
\DeclareSIUnit{\nothing}{\relax}
\usepackage{hyperref}
\usepackage{multirow}
\usepackage[nolist,nohyperlinks]{acronym}

\usepackage{booktabs}
\usepackage[nameinlink, capitalise]{cleveref}
\usepackage{svg}
\usepackage{tikz}
\usetikzlibrary{automata, positioning, arrows}
\usepackage{enumerate}
\usepackage[inline]{enumitem}
\makeatletter
\newcommand*{\org@overidelabel}{}
\let\org@overridelabel\@verridelabel
\@ifpackagelater{acronym}{2015/03/21}{
  \renewcommand*{\@verridelabel}[1]{%
    \@bsphack
    \protected@write\@auxout{}{\string\AC@undonewlabel{#1@cref}}%
    \org@overridelabel{#1}%
    \@esphack
  }%
}{
  \renewcommand*{\@verridelabel}[1]{%
    \@bsphack
    \protected@write\@auxout{}{\string\undonewlabel{#1@cref}}%
    \org@overridelabel{#1}%
    \@esphack
  }%
}
\makeatother

\acmYear{2023}\copyrightyear{2023}
\setcopyright{acmlicensed}
\acmConference[ENSsys' 23]{11th International Workshop on Energy Harvesting \& Energy-Neutral Sensing Systems}{November 12, 2023}{Istanbul, Turkiye}
\acmBooktitle{11th International Workshop on Energy Harvesting \& Energy-Neutral Sensing Systems (ENSsys' 23), November 12, 2023, Istanbul, Turkiye}
\acmPrice{15.00}
\acmDOI{10.1145/3628353.3628545}
\acmISBN{979-8-4007-0438-3/23/11}

\usepackage{eso-pic}
\usepackage{url}
\AddToShipoutPictureBG*{
  \AtPageUpperLeft{%
    \put(0,-40){\raisebox{15pt}{\makebox[\paperwidth]{\begin{minipage}{21cm}\centering
      \textcolor{gray}{This article has been accepted for publication in the\\Proceedings of the 11th International Workshop on Energy Harvesting and Energy-Neutral Sensing Systems (ENSSys '23).\\
      DOI: 10.1145/3628353.3628545}
    \end{minipage}}}}%
  }
  \AtPageLowerLeft{%
    \raisebox{25pt}{\makebox[\paperwidth]{\begin{minipage}{21cm}\centering
      \textcolor{gray}{ \copyright 2023 Authors and ACM. This is the author's version of the work. It is posted here for your personal use. Not for redistribution. The definitive Version of Record will be published in \textit{Proceedings of the 11th International Workshop on Energy Harvesting and Energy-Neutral Sensing Systems}, https://doi.org/10.1145/3628353.3628545.
      }
    \end{minipage}}}%
  }
}

\begin{document}

\title{Energy-Aware Adaptive Sampling for Self-Sustainability in Resource-Constrained IoT Devices}

\author{Marco Giordano}
\authornote{Both authors contributed equally to the paper}
\orcid{0000-0003-1106-3472}
\email{marco.giordano@pbl.ee.ethz.ch}
\affiliation{%
  \institution{D-ITET, ETH Zürich}
  \city{Zürich}
  \country{Switzerland}
}

\author{Silvano Cortesi}
\authornotemark[1]
\orcid{0000-0002-2642-0797}
\email{silvano.cortesi@pbl.ee.ethz.ch}
\affiliation{%
  \institution{D-ITET, ETH Zürich}
  \city{Zürich}
  \country{Switzerland}
}

\author{Prodromos-Vasileios Mekikis}
\orcid{0000-0003-0616-0657}
\email{akis.mekikis@hilti.com}
\affiliation{%
  \institution{Hilti Corporation}
  \city{Schaan}
  \country{Liechtenstein}
}

\author{Michele Crabolu}
\orcid{0000-0003-4308-2432}
\email{michele.crabolu@hilti.com}
\affiliation{%
  \institution{Hilti Corporation}
  \city{Schaan}
  \country{Liechtenstein}
}

\author{Giovanni Bellusci}
\orcid{0000-0002-5514-2503}
\email{giovanni.bellusci@hilti.com}
\affiliation{%
  \institution{Hilti Corporation}
  \city{Schaan}
  \country{Liechtenstein}
}

\author{Michele Magno}
\orcid{0000-0003-0368-8923}
\email{michele.magno@pbl.ee.ethz.ch}
\affiliation{%
  \institution{D-ITET, ETH Zürich}
  \city{Zürich}
  \country{Switzerland}
}

\begin{abstract}
In the ever-growing \ac{IoT} landscape, smart power management algorithms combined with energy harvesting solutions are crucial to obtain self-sustainability. This paper presents an energy-aware adaptive sampling rate algorithm designed for embedded deployment in resource-constrained, battery-powered \ac{IoT} devices. The algorithm, based on a \ac{FSM} and inspired by \ac{TCP} \textsc{Reno}'s additive increase and multiplicative decrease, maximizes sensor sampling rates, ensuring power self-sustainability without risking battery depletion. 
Moreover, we characterized our solar cell with data acquired over 48 days and used the model created to obtain energy data from an open-source world-wide dataset.

To validate our approach, we introduce the \textsc{EcoTrack} device, a versatile device with \ac{GNSS} capabilities and \ac{LTE-M} connectivity, supporting \textsc{MQTT} protocol for cloud data relay. This multi-purpose device can be used, for instance, as a health and safety wearable, remote hazard monitoring system, or as a global asset tracker.

The results, validated on data from three different European cities, show that the proposed algorithm enables self-sustainability while maximizing sampled locations per day. In experiments conducted with a \qty{3000}{\milli\ampere{}\hour} battery capacity, the algorithm consistently maintained a minimum of 24 localizations per day and achieved peaks of up to 3000.

\acresetall
\end{abstract}

\ccsdesc[500]{Hardware~Power estimation and optimization}
\ccsdesc[500]{Computing methodologies~Optimization algorithms}
\ccsdesc[300]{Hardware~Sensors and actuators}
\ccsdesc[500]{Networks~Location based services}

\keywords{sensor network, embedded systems, low-power, sustainability,energy harvesting, adaptive sampling rate, GNSS, LTE, tracking, bluetooth low energy}

\maketitle

\section{Introduction}
\label{sec:introduction}

Estimates suggest that there will be a deployment of approximately 30 billion \ac{IoT} devices by the year 2025~\cite{rawat2022recent, islam23_amalg_inter_comput_system}. Such a number of connected devices poses a significant problem regarding energy sourcing: if all these devices were battery-powered, it would require a tremendous effort to recharge/change batteries, and it would impose a serious toll on emissions, cost, and overall sustainability~\cite{weddell13_survy_multi_sourc_energ_harves_system, mileiko23_statef_energ_manag_multi_sourc}.

Energy harvesting presents a promising solution to this problem, as it offers the potential to extract electrical power from various environmental sources\cite{murphy15_big}. These include solar irradiation, thermal gradients, radio frequency, tribology, and kinetic energy, among others. However, scavenging ambient power is unreliable and inconsistent in nature~\cite{bader10_enabl_batter_less_wirel_sensor}: sources like solar energy vary with the period of the year, time of the day, and geographical location. Rechargeable batteries and super-capacitors serve as storage elements to cope with such fluctuations, but are often limited in capacity by size and cost constraints, potentially leading to system failure. Therefore, meticulous energy resource management is required, and smart energy-aware algorithms are a very promising avenue\cite{magno09_adapt,giordano23_optim_iot_based_asset_utiliz_track}.

Smart energy-aware algorithms can contribute to ensuring system operation is adapted according to the available energy~\cite{srbinovski15_energ, mayer22_model_based_design, balsamo16_hibernus}. 
Literature provides several potential solutions in this field, such as \ac{DVFS}, energy-aware scheduling algorithms~\cite{sabovic2021demonstration}, and wireless sensor networks protocols~\cite{mittal2019survey}. However, oftentimes the activity with the highest power consumption is sensing, and therefore a low-power self-sustaining \ac{IoT} sensor node must adapt its sampling frequency to accommodate energy requirements~\cite{giouroukis2020survey}. Adaptive sampling is a well-known technique, which found use also in energy harvesting systems~\cite{heo22_adaptivesampling, buchli15_optimal_power_management}. However, many of the currently proposed implementation requires complex algorithm, with hefty computational and power requirements.

This paper proposes a \ac{FSM} based algorithm that aims at maximizing a certain utility function, i.e. the sampling rate of a sensing system, given a specific battery size. 
In particular, our algorithm uses only the battery state of charge, as a proxy for the harvested energy, and is, therefore, agnostic to the energy source and eventual seasonality.  We compare our algorithm to the state-of-the-art \ac{FHC}, showing that the simplistic design of the algorithm minimizes computational cost and memory footprint while marginally losing on performance. 

To validate the proposed algorithm, this paper introduces \textsc{EcoTrack}, a platform designed with low power and energy efficiency in mind. It can provide worldwide localization in an energy-neutral fashion, featuring \ac{GNSS} and \ac{LTE-M} connectivity to enable localization and long-range gateway-less communication. Being always connected to the internet, the system is also continuously listening to alarms sent by the cloud to signal, for example, a hazardous condition on the work site.

Moreover, the device has a flexible nature: it can be repurposed as an asset tracking device with just a firmware change, as further analyzed and reported in this work, or deployed in potentially harmful environments to track forest fires, requiring no human intervention for sustained operation.

A thorough evaluation of power requirements and a breakdown of power consumption is reported, as well as a simulation of self-sustainability at different latitudes, which showed failure-free operation over the time span of two years. Finally, the implementation of the developed algorithm, both in C and Python, has been made publicly available\footnote{\url{https://github.com/ETH-PBL/EcoTrack}}.

The main contributions of this paper are listed as follows:
\begin{itemize}
    \item The design and implementation of an energy-aware adaptive sampling rate algorithm based on an \ac{FSM} in order to achieve energy-neutrality in resource-constrained battery-powered systems.
    \item The design of a low power \ac{GNSS}, \ac{LTE-M} and \ac{BLE} enabled sensor node.
    \item The creation of a global solar irradiation model based on real-world data.
    \item The evaluation of the proposed algorithm on the prototype system introduced in this paper with a comparison to the state of the art.
\end{itemize}

The paper is organized as follows: \cref{sec:related_work} surveys related work in adaptive algorithms, with a focus on energy harvesting, worker safety, and asset tracking devices. 
\cref{sec:algo_design} explains the proposed energy harvesting solution to achieve zero-energy operation. \cref{sec:validation} provides an overview of the system and each building block is explained thoroughly, motivating the choices in terms of performance and power consumption. \cref{sec:results} gives a detailed power analysis performed on the developed \textsc{EcoTrack}, showcasing our developed adaptive sampling algorithm in three cities at different latitudes. \cref{sec:conclusion} draws conclusions from this work and future work is laid down.
\vspace{-0.2cm}
\section{Related Work}
\label{sec:related_work}
Previous works have demonstrated the key role of energy-aware algorithms in achieving self-sustainability~\cite{mayer22_model_based_design}.

In \cite{mekikis18_connectivity}, the authors investigate the connectivity performance of a large-scale sensor network, i.e., the ability of all sensors to deliver their data, that harvests solar energy to achieve a self-sustainable zero-energy network. Specifically, they propose an energy-aware algorithm that utilizes a solar radiation and cloud-cover model of a specific location, to adjust the frequency and power of the sensor transmissions. Their analytical framework provides closed-form solutions for the sensor transmission probability and the end-to-end connectivity probability, which can be employed to determine the parameters and conditions that ensure a self-sustainable network operation throughout the year.

In~\cite{moser_adaptive_power_management}, Moser et al. implemented an optimal \ac{MPC} based algorithm, which adaptively manages the power of a self-sustaining system.
Their system model captures the incoming energy, the storage capacity as well as tasks of various system components. As the incoming energy is unknown a priori, a predictor estimates it on a finite horizon in the first phase. Although more complex predictors can be used, for their evaluation they rely on the periodicity of the solar power (i.e. over each day) and knowledge of the already harvested energy over the past days. Using a cost function that maximizes the execution rate of the tasks, an \ac{MPC} is then used to estimate the future execution rates over an entire horizon. By limiting the horizon and optimizing the memory footprint it was possible to implement the algorithm online on an \textsc{ATmega128L}. The execution time of the algorithm itself is around \qty{2}{\milli\second}, which corresponds to \qty{32}{\kilo\nothing} operations on their platform, over \(40\times\) more than our implementation (architecture compensated).

Instead of relying on an \ac{MPC}, Thiele et al. employed a \acf{FHC} in~\cite{buchli15_optimal_power_management, ahmed19_optimal_power_manag} and~\cite{draskovic21_optimal_power}, respectively. By assuming that the annual harvestable energy does not change much, the sampling rate over the horizon of a whole year is optimized using knowledge of the incoming energy, usually obtained from a predictor. Using the boundary condition that the battery state must be the same at the beginning and the end of the horizon and that the battery never runs out, an optimal strategy can be planned. However, it should be emphasized that this algorithm needs perfect knowledge of the incoming energy of the coming year to achieve optimality. For online implementations, this results in requiring a predictor of the future incoming energy and thus will not lead to an optimal control. The error with respect to the optimal control highly depends on the used predictor.
In~\cite{draskovic21_optimal_power}, Draskovic et al. implemented the \ac{FHC} on a \textsc{TI MSP432} \ac{MCU} using a simple \ac{EWMA} on past data as predictor. Due to the iterative solver that such an algorithm requires, solving the optimization problem over a horizon of \qty{8}{\hour} took \qty{0.2}{\second}, about \(\qty{25}{\kilo\nothing}\) times slower than our implementation.

In~\cite{heo22_adaptivesampling}, Heo et al. implement a predictive energy-aware adaptive sampling method for \acp{WSN}. Based on historic recordings of the harvested energy as well as the current battery level, a deep actor-critic reinforcement learning implementation is used to adapt the sampling rate accordingly. The implementation consists of three states: Up, Stay, and Down. Based on two different agents, one being feed-forward and the other recurrent, probabilities for the next actions are evaluated. The implementation itself is built on \textsc{OpenAI Gym}~\cite{openai_gym} and is therefore not (yet) suited for a low-power embedded application. 

Although these algorithms often achieve optimality, they have several limitations. While all of them are computationally expensive, \ac{MPC}- and \ac{FHC}-based approaches rely on a predictor of the incoming power, which adds another layer of complexity to the algorithm. In contrast, the approach proposed in this paper is inspired by the congestion control used in the networking communication field, and relies on a simple \ac{FSM}-based approach building on the core concepts of \ac{TCP} congestion control, thus resulting in a simpler and more suited solution for \ac{IoT} devices. Congestion has been identified as a networking problem in 1984~\cite{rfc896}, and congestion control has ever since been researched, with the most prominent algorithm being \ac{TCP} \textsc{Reno}~\cite{rfc5681}. Although \ac{TCP} \textsc{Reno} has already been applied for adaptive sampling in \cite{alhoqani15_asa, monteiro17_asa}, to the best of our knowledge there is no existing energy-aware algorithm exploiting \ac{TCP} \textsc{Reno} that maximizes the sampling rate with respect to available energy.\vspace{-0.2cm}
\section{Algorithm Design}
\label{sec:algo_design}
To optimally use the harvested energy and avoid failure states, the sampling rate of power-hungry sensors needs to be adapted: if the rate is too high, the battery will empty during the weakest days of the year (or it requires to be much larger), while if the rate is too low, localizations are lost during days with more energy available.  In our use case, the localization rate of the \ac{GNSS} module has to be adapted in order to achieve the highest possible localization rate without running out of power.
Therefore, in the following, we present a low-complexity and energy-aware algorithm to change the localization rate online, depending on the available energy in order to have the optimal trade-off performance vs lifetime. 

\subsection{Adaptive Sampling Algorithm}
To optimize the adaptive sampling rate depending on the incoming energy, using the battery's state of charge as a proxy, the paper proposes an \ac{FSM}-based algorithm. Inspired by congestion control in network protocols, we identified the network throughput is equivalent to the localization samples per day, and the congestion of the network is equivalent to the battery state - which in turn depends on external influences, in our case on the total net income of energy. We, therefore, decided to implement its \ac{AIMD} algorithm, with a decrease factor of 2 as in \ac{TCP} \textsc{Reno} and an increase rate of one measurement per hour (see \cref{fig:statemachine}). The state change between increasing or decreasing the rate is determined by an \ac{FSM} (\cref{subsubsec:fsm}) and a metric function (\cref{subsubsec:metric}).

\subsubsection{\acf{FSM}}\label{subsubsec:fsm}
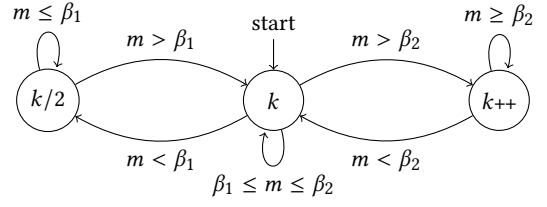
\begin{figure}[htpb!]
    \centering
    \vspace{-0.2cm}
    \begin{tikzpicture}
[align=center,node distance=3cm,
state/.style={circle, draw, minimum size=0.8cm}]
\node[state] (q1) {$k/2$};
\node[state, initial, initial where=above, right of=q1] (q2) {$k$};
\node[state, right of=q2] (q3) {$k{\scriptstyle++}$};
\draw [->] (q1) edge[loop above] node{$m\leq \beta_1$} (q1)
(q2) edge[loop below] node{$\beta_1\leq m\leq \beta_2$} (q2)
(q3) edge[loop above] node{$m\geq \beta_2$} (q3)
(q1) edge[bend left, above] node{$m>\beta_1$} (q2)
(q2) edge[bend left, below] node{$m<\beta_1$} (q1)
(q2) edge[bend left, above] node{$m>\beta_2$} (q3)
(q3) edge[bend left, below] node{$m<\beta_2$} (q2);
\end{tikzpicture}
    \vspace{-0.2cm}
    \caption{State diagram of the \acs{FSM}-based adaptive sampling algorithm: \(k/2\) means that \(k\) halves in the next step; \(k++\) means that \(k\) increases by a fixed rate in the next step, in our case 24; \(k\) means that \(k\) stays the same in the next step.}
    \label{fig:statemachine}
    \vspace{-0.3cm}
\end{figure}

The \ac{FSM} of the algorithm is represented in \cref{fig:statemachine} with \(k\) being the number of sensor samplings per day. 
Each state shows whether the number of samplings per day, \(k\), remains constant or is altered with respect to its old value. The parameters \(\beta_1\) and \(\beta_2\) determine together with the metric function \(m\) the threshold levels for state changes. Together with \(\gamma\), the battery level threshold always leads to a multiplicative increase, they are the three parameters that need to be tuned.

To ensure feasibility under all circumstances, the optimization of the algorithm must be done under the worst-case conditions. This is for the reason that the algorithm does not rely on a predictor of the incoming energy and thus can not react to conditions worse than it was optimized on.

\subsubsection{Metric Function}\label{subsubsec:metric}
In \cref{eq:metric}, we outline the metric function \(m\). Based on this function, which considers the change in battery capacity over the previous day, the algorithm adjusts the number of daily samplings, performing more or fewer as needed.

\begin{equation}\label{eq:metric}
    m = \underbrace{\vphantom{\begin{cases}0\\0\end{cases}} B\cdot(b[t] - b[t-1])}_{\text{batt. state difference}} - \underbrace{\vphantom{\begin{cases}0\\0\end{cases}} \left(\frac{1}{b[t]}-1\right)}_{\text{low batt. penalty}} + \underbrace{\begin{cases}
    \infty,& \text{if } b[t] \geq \gamma\\
    0,     & \text{otherwise}
\end{cases}}_{\text{high batt. reward}}
\end{equation}

where \(b[t]\) is the battery charge in percentage at a time \(t\) (\(t\) in days), \(B\) the total capacity of the battery, and \(\gamma\) is the boundary for the battery charge, above which the algorithm increases the localization rate. In general, the metric \(m\) gets rewarded when
\begin{enumerate*}[label=(\roman*),,font=\itshape]
    \item the battery's charge is high;
    \item the battery's charge increased;
\end{enumerate*}
and punished when
\begin{enumerate*}[label=(\roman*),,font=\itshape]
    \item the battery's charge is low;
    \item the battery's charge decreased.
\end{enumerate*}

\subsubsection{Algorithm Tuning}\label{subsubsec:algo_tuning}
To find the optimal parameters for \((\beta_1,\,\beta_2, \gamma)\) of the state-machine and the metric function, the three parameters have to be found by solving the optimization problem \cref{eq:costfct} under the following inequality constraints:
\begin{align}\label{eq:constraints}
\begin{split}
    b[\cdot] &\geq 5\%\\
    b[\cdot] &\leq 100\%\\
    k &\geq 24\\
\end{split}
\end{align}
The overall goal of the parameter search is the maximization of \(k\) over the course of a full year (\cref{eq:costfct}) while ensuring at least one positioning per hour.
\begin{equation}\label{eq:costfct}
    J = \max_{k\in\mathbb{N}^+}\left(\sum\limits_{t=0}^T k[t]\right)\ \text{subj. to \cref{eq:constraints}}
\end{equation}
with \(T\) being a full year and \(k\) evolving as described in \cref{subsubsec:fsm}.\vspace{-0.2cm}
\section{Materials and Methods}
\label{sec:validation}
In this section, we introduce the platform employed to test the algorithm, including the system design choices for optimal power management and a detailed power consumption breakdown of the prototype. The setup and data collection used to validate the algorithm are also reported.

\subsection{Validation Platform}
\label{subsec:system_design}

\begin{figure}[htpb!]
    \centering
    \includesvg[width=0.9\columnwidth]{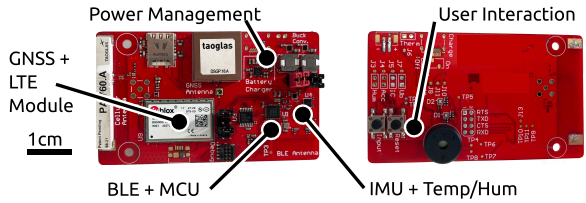}
    \vspace{-0.3cm}
    \caption{Overview of the developed \textsc{EcoTrack}}
    \label{fig:system_overview}
    \vspace{-0.3cm}
\end{figure}

\subsubsection{Hardware}
The components of the \textsc{EcoTrack} were carefully selected as a trade-off between performance and availability of low-power states, targeting an online-adaptable duty-cycled application.

At the heart of the badge, there is the \ac{GNSS} and \ac{LTE-M} module. The \textsc{SARA R510M8S} from \textsc{u-blox} has been selected as it combines the \textsc{SARA R510S} \ac{LTE-M} module with the \textsc{M8} \ac{GNSS} receiver, which can be run simultaneously and independently, offering a sweet spot design in terms of physical space and power. The module also hosts a hardware-based root of trust, which is essential for providing security features. In our application, it has been exploited to provide a secure \textsc{MQTT} connection to \ac{AWS} cloud, asynchronously encrypting every message exchanged by the badge and the servers.
Moreover, the \ac{LTE-M} module allows to download \ac{GNSS} assistance packets to improve energy efficiency and time to fix. In particular, these packets include:
\begin{enumerate*}[label=(\roman*),,font=\itshape]
    \item ephemeris, giving up-to-date information about the satellites' health;
    \item clock correction coefficients and orbital information (valid for few hours); and 
    \item almanac data, providing satellites' orbits (up to 90 days).
\end{enumerate*}

The \ac{MCU} of choice is the \textsc{nRF52833}, a Cortex-M4 from \textsc{Nordic Semiconductor} operating at a frequency of \qty{64}{\mega\hertz} and offering \qty{512}{\kilo\byte} of flash and \qty{128}{\kilo\byte} of \acs{RAM}. It features a power consumption as low as \qty{52}{\micro\ampere/\mega\hertz} and provides a low-power state with full RAM retention at just \qty{1.8}{\micro\ampere}.

The badge also features a 6-DoF \ac{IMU} from \textsc{STMicroelectronics}, the \textsc{LSM6DSLTR}. It offers a high sample rate of up to \qty{6664}{\hertz} for both accelerometer and gyroscope, with a full scale of \(\pm \qty{8}{G}\) and \(\pm \qty{2000}{dps}\) respectively. It can achieve low power operation down to \qty{9}{\micro\ampere} for accelerometer-only continuous sampling at \qty{12.5}{\hertz}, which allows for motion and drop detection using the chip's built-in functions.

Finally, the badge hosts the \textsc{SHTC3}, a temperature and humidity sensor from \textsc{Sensirion}, and a series of accessory electronics, such as a fuel gauge, a battery charger, different power converters, and load switches for optimal power management and minimal leakage.

\subsubsection{Firmware}
Given the complexity of the firmware and all the subsystems involved, the firmware framework was designed based on \textsc{Zephyr}, an open-source and cross-platform \ac{RTOS}.

\begin{figure}[htpb!]
    \centering
    \includegraphics[width=0.8\columnwidth]{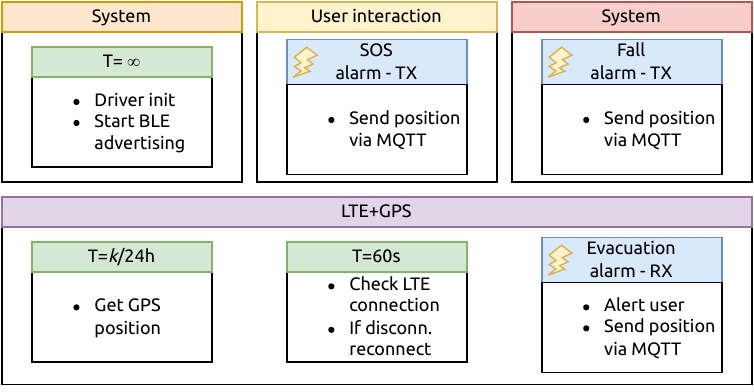}
    \vspace{-0.1cm}
    \caption{RTOS task overview. The lightning represents an interrupt-enabled thread.}
    \label{fig:firmware_overview}
    \vspace{-0.4cm}
\end{figure}

An overview of the firmware structure is provided in \cref{fig:firmware_overview}. The application starts initializing the drivers for the sensors and the \textsc{SARA}, proceeds then to start advertising the \ac{BLE}, and then the init thread goes indefinitely into sleep. One of the threads checks whether the module is connected, and, in case it is not, tries to register again to the network. Another thread handles the localization and can either run periodically or on-demand if signaled with a semaphore. 

The last two logically important threads are the alarm and \ac{IMU} wake-up thread. Those threads can be woken up by one of the following active interrupts on the badge:
\begin{enumerate*}[label=(\roman*),,font=\itshape]
    \item an SOS button; 
    \item the free fall interrupt from the \ac{IMU}, or 
    \item an \textsc{MQTT} RX message on the Evacuation topic.
\end{enumerate*}
The first can be triggered by the user in case they witness a dangerous situation, the fall alarm is triggered by the \ac{IMU} in case a fall is detected, and the last is a message sent to all the workers in a particular zone to evacuate after the danger has been reported.

\subsubsection{Power Estimation}
\label{subsubsec:power_estimation}

\begin{table}[htpb!]
\caption{Power consumption breakdown of the different \textsc{EcoTrack} status. The system voltage is \qty{3.8}{\volt}}
\label{tab:power_profilign}
\vspace{-0.2cm}
\begin{tabular}{@{}llr@{}}
\toprule
\textbf{Component} & \textbf{Power state} & \textbf{Power consumption} \\ \midrule
Baseline  &               Switched-off &  \qty{0.296}{\micro\watt}   \\ \midrule
\multirow{2}{*}{Baseline - \acs{MCU}} & \acs{BLE} adv. (\acs{UART} on)                  & \qty{4.94}{\milli\watt}                                   \\
    &               \acs{BLE} adv. &  \qty{885.0}{\micro\watt}   \\
                                \hline
\multirow{6}{*}{\textsc{SARA}}           & \acs{GPS}                              &    \qty{258.40}{\milli\watt}                                   \\
                                & Idle Connected             &    \qty{9.12}{\milli\watt}                                    \\
                                & \acs{LTE} connecting                              &    \qty{456.00}{\milli\watt}                                    \\
                                & Idle                             &    \qty{2.66}{\milli\watt}                                    \\
                                & Deep sleep                       &    \qty{266.0}{\micro\watt}   \\
                                & Energy per localization & \qty{4.7}{\joule}\\
                                \midrule
\multirow{1}{*}{\acs{IMU}}
                                & Free fall / wake                 &    \qty{760.00}{\micro\watt}                                    \\
                                \hline
\multirow{2}{*}{Temperature}    & Measure                          &    \qty{1.90}{\milli\watt}                                    \\
                                & Sleep                            &    \qty{190.00}{\micro\watt}                                \\
\midrule
\midrule
\multirow{2}{*}{\textbf{Summary}}    & \textbf{Localization}                          &    \textbf{\qty{5.1}{\joule}}                                    \\
                                & \textbf{Low power idling}                            &    \textbf{\qty{19}{\milli\watt}}                                \\

                                \bottomrule
\end{tabular}
\vspace{-0.2cm}
\end{table}

A thorough power profiling was performed on the developed smart node, with special attention towards the \ac{GPS} and \ac{LTE-M} modules. In \cref{tab:power_profilign}, we report a power consumption breakdown for each component in the \textsc{EcoTrack} in different power states.
Starting with \textsc{EcoTrack} connected to a battery but still in a powered-off state, the system switch has a sub-\unit{\micro\watt} leakage from the system switch.
The MCU has two different baselines depending on whether the \acs{UART} is always listening (necessary to receive messages from the \textsc{SARA} module), or in an OFF state. In both cases, the MCU is in low power when not active and advertising BLE with a period of \qty{7}{\second}.
The most power-hungry component in active mode, as expected, is the \textsc{SARA} module, with the peak power during the \ac{LTE-M} connection and \ac{GPS} tracking. Staying always connected consumes significantly more power 

Complementing the power profiling, there are two more sensors, the \ac{IMU} and the temperature and humidity. The former is continuously sampling at \qty{12.5}{\hertz} with onboard always-on fall detection activated, while the second is wakened up in a configurable interval to collect a temperature and humidity sample.

Having both \ac{GNSS} and \ac{LTE} in the same package allows for enhancing the \ac{GNSS} performance by downloading the location information from an online server rather than directly from the satellite. Assisted \ac{GNSS} can obtain fixes in seconds, up by two orders of magnitude faster than \ac{GNSS}-only, especially with first position fixes. In this work, we choose to use only the \ac{GNSS} (without any aid) because it does not require any subscription and represents the worst case in terms of energy consumption.

\subsection{Solar Harvesting and Dataset}\label{subsec:solar_dataset}
To guarantee the necessary average power, as well as the resulting energy costs per localization, the device was equipped with two \textsc{SM141K08TFV} solar cells from \textsc{AnySolar} connected in parallel with a total surface of \qtyproduct{45 x 58}{\milli\meter}, and a \textsc{BQ25570} Energy Harvester from \textsc{TI}. As the solar cells' datasheet does not provide the power-vs-irradiance characteristic, in a first step the harvesting circuit including the solar cells had to be characterized. Then, in order to prove that our adaptive sampling algorithm is robust to different latitudes, i.e., places with different solar irradiation potential, solar irradiance datasets from different latitudes were fed to our model to compute the harvested power.

\begin{figure*}\centering
    \vspace{-0.2cm}
    \includegraphics[width=0.92\textwidth]{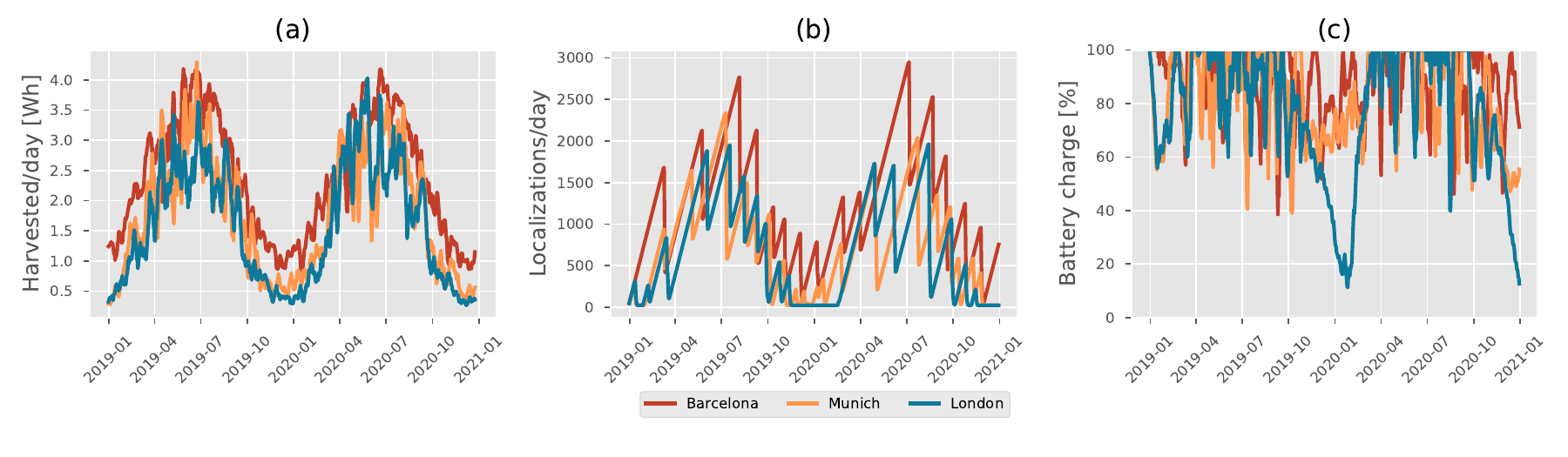}
    \vspace{-0.5cm}
    \caption{\acs{FSM}-based adaptive sampling algorithm applied on the validation data from Barcelona, Munich, and London.\\
    (a) Harvested energy per day. (b) Localizations \(k\) per day. (c) Evolution of the battery's charge.}
    \label{fig:energyharvesting}
    \vspace{-0.2cm}
\end{figure*}

\subsubsection{Solar Cell Model}
Using a \ac{SMU} (\textsc{Keysight B2902A}) and an ambient light sensor (\textsc{TI OPT3002}) the solar cells and harvester were exposed to daylight for 48 days and the solar irradiance and harvested power were measured. Thereby, the SMU simulated a battery with a voltage of \qty{3.7}{\volt} as variable load.

As the \textsc{OPT3002} saturates at an irradiance of \qty{100}{\watt\per\meter\squared}, the linear relation between harvested power and solar irradiance~\cite{miozzo14_solarstat, gianfreda_solarstat, politaki17_stochastic_model} has been exploited to obtain a model between harvested power and solar irradiance:
A Ridge Regression as in \cref{eq:ridgereg} has been applied to a randomly sampled train-test split of our (non-saturated) dataset (\(80\%/20\%\)).
\begin{equation}\label{eq:ridgereg}
f:\ E_e \to P,\ \alpha=0.1
\end{equation}
with \(E_e\) being the irradiance (\unit{\watt\per\meter\squared}) and \(P\) the harvested power (\unit{\watt}).
The \ac{RMSE} of the obtained model when predicting the harvested power using the solar irradiance of the test set was \qty{4.81}{\milli\watt}, while the \ac{MAE} was \qty{2.20}{\milli\watt} which corresponds to an error of less than 0.5\% with respect to the maximum peak power of the solar cell.

\subsubsection{Dataset Application}
The next step was to apply the model of our characterized system to the historical data from the \textsc{SARAH-2} dataset~\cite{sarah2} to determine the harvestable energy at different locations around the world over an entire year. The irradiance data of the vertical axis at a \ang{0} slope of the three European cities Berlin, Rome, and Zurich, which are located at different latitudes, were taken for tuning of the algorithm's parameters, those of London, Munich, and Barcelona for validating its functionality.

When looking at the statistical measures of the estimated harvested power of the three cities, we find that the maximal harvested power, especially when considering only single days, is approximately \(\qty{183}{\milli\watt}\), which is similar across the three cities. On the other hand, the minimal harvested power differs greatly across those cities: while all of the cities had their weakest days in December or January, it can be seen that the further north a city is located, the less solar irradiance it receives, resulting in a decrease in the amount of power that can be harvested. The minimal average power over one day with \qty{3.73}{\milli\watt} is collected in Berlin, whereas the least sunny day in Rome gave an average power of \qty{9.94}{\milli\watt}.

Using the data of the training cities together with the device's power consumption as in \cref{tab:power_profilign} (highlighted in bold), the algorithm's parameters have been determined as described in \cref{subsubsec:algo_tuning}. Specifically, the final parameters have to be valid individually for all the locations. To that end, the optimal parameters have been determined to be \(\beta_1=-0.203\), \(\beta_2=0.468\) and \(\gamma=0.67\).\vspace{-0.2cm}

\begin{figure}[htpb!]
    \centering
    \includegraphics[width=0.8\columnwidth]{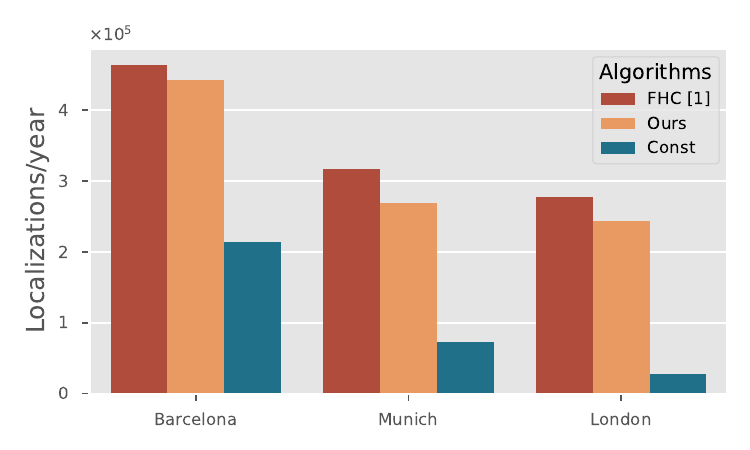}
    \vspace{-0.5cm}
    \caption{Comparison in total locations per year of the \ac{FHC} algorithm, our algorithm, and a constant sampling rate.}  
    \label{fig:algo_comparison}
    \vspace{-0.3cm}
\end{figure}
\section{Results}
\label{sec:results}

In this section, we validate the results of the determined parameters using the solar data of three new locations: London (northern Europe), Munich (central Europe), and Barcelona (southern Europe).
\cref{fig:energyharvesting} illustrates the data collected from the validation set of cities. More specifically, \cref{fig:energyharvesting}(a) shows the harvested energy, accumulated per day. As expected, Barcelona is the city with the highest harvested energy, while London has the least, and Munich is a middle-ground among the three. This trend is also noticeable in Figure~\ref{fig:energyharvesting}(b), where the algorithms' results are reported in terms of localizations per day. The well-known "TCP sawtooth" pattern is recognizable in the plot. Lastly, in Figure~\ref{fig:energyharvesting}(c), the battery state of charge is plotted. Even though the parameters were obtained in other locations, we can observe that the battery charge of the devices never drops below 10\%. Notable is, that for London, where there is substantially less energy during winter than in the other cities, the battery drops close to 10\%. This is expected, as its harvested energy is close to the one that determined the tuning parameters. Moreover, the algorithm manages to successfully adapt the rate of localization when more energy is available, i.e., as the amount of harvested energy rises or falls, there is a corresponding increase or decrease in the localization rate, respectively.

To underline the efficiency of our algorithm, we compared it to the \ac{FHC} algorithm from~\cite{ahmed19_optimal_power_manag} as well as a constant sampling rate. The constant rate of localizations was optimized individually on the historical data for each of the three cities London, Munich, and Barcelona, so that the battery charge never falls below 5\%. The \ac{FHC} algorithm was run with the same boundary conditions as our algorithm, except that it was given the full history data as a predictor, in a clairvoyant setting. It optimizes mathematically the number of localizations per hour and represents the maximum possible with the given boundary conditions and the condition \(b[0] = b[T]\), in a horizon of 365 days. As it is clairvoyant by knowing already the incoming energy, it is in reality not implementable and its result shows how close one can come to the optimum.

A comparison of the two algorithms together with a constant localization rate (tuned to each city individually) can be seen in \cref{fig:algo_comparison}. The total number of locations per year is reported, and, as expected, the constant rate of localizations performs the poorest because many more measurements would be possible in summer, whereas the optimal \ac{FHC} performs the best. Our algorithm performs too conservatively in London ($0.88\times$ of FHC) and Munich ($0.85\times$ of FHC) and is closer to optimality in Barcelona, where there is more energy available ($0.95\times$ of FHC). However, compared with the constant localization rate, we achieved an increase of $2.06\times$ and $8.68\times$ more localization per year in Barcelona and London, respectively.  For fairness, we kept the same tuning on the different cities to evaluate it in fair condition, on the other hand, knowing the energy expectation according to the reference city our algorithm can be tuned to reach a higher localization rate when needed. Moreover, this evaluation does not include the energy overhead due to the adaptive algorithm as we wanted to show the theoretical limits of the 3 algorithms without the platform-dependent energy overhead, which is analyzed below.

To evaluate the energy and computational overhead of both our algorithm and the \ac{FHC} we proceed as follows: we implemented and profiled our adaptive sampling algorithm on the \ac{MCU} of the \textsc{EcoTrack} (\textsc{ARM Cortex-M4F}-based \textsc{nRF52833}, clocked at \qty{64}{\mega\hertz}), resulting in only 527 cycles per execution. Regarding the \ac{FHC}, the authors in~\cite{draskovic21_optimal_power} did not state the clock frequency of their MCU, therefore we evaluated numbers from \qty{3}{\mega\hertz} (default clock frequency~\cite[p.~14]{msp432}) up to \qty{48}{\mega\hertz} (maximum frequency) to be the \ac{MCU} clock speed, making their embedded implementation requiring \qty{0.6}{\mega\nothing} cycles in the best case, as shown in \cref{tab:comparisonalgorithms}. It is important to notice that both \acp{MCU}, the \textsc{MSP432} and the \textsc{nRF52833}, use the same \textsc{ARM Cortex-M4F}-based core, so the comparison is done in fair conditions and using the default clock frequency leads to the best possible clock count and energy consumption for the \ac{FHC}. Additionally, our implementation needs to be executed only once per day and requires just knowledge of the current battery state and the one of the day before. On the other hand, the embedded \ac{FHC} implementation has to be executed every hour and needs additional hardware to sense the harvested energy and more memory to store recorded data over a full horizon of \qty{8}{\hour}. Its memory footprint scales therefore with a complexity of \(\mathcal{O}(n)\), since it has to store past values over a full horizon to run a prediction. Ours needs only the previous state of the battery, therefore presenting a memory complexity of \(\mathcal{O}(1)\). 

\begin{table}[htpb!]
    \vspace{-0.3cm}
    \caption{Computational cost comparison between state-of-the-art adaptive sampling algorithms. The cycle cost of~\cite{draskovic21_optimal_power} is estimated using frequency from [\qty{3}{\mega\hertz}, \qty{48}{\mega\hertz}] of the \acs{MCU}.}
    \label{tab:comparisonalgorithms}
    \vspace{-0.2cm}
    \centering
    \begin{adjustbox}{max width=\columnwidth}
    \begin{tabular}{@{}llrr@{}}
    \toprule
        \textbf{Algorithm} & & \textbf{Duration} & \textbf{Computational cost}\\
    \midrule
        \acs{FHC}~\cite{draskovic21_optimal_power} & \textsc{MSP432} & \qty{200}{\milli\second} & \(\approx\qty{0.6}{\mega\nothing}-\qty{9.6}{\mega\nothing}\) cycles\\
        static rate & \textsc{nRF52833} & \qty{0}{\micro\second} & \(0\) cycles\\
        this work & \textsc{nRF52833} & \qty{8.24}{\micro\second} & \(527\) cycles\\
    \bottomrule
    \end{tabular}
    \end{adjustbox}
    \vspace{-0.2cm}
\end{table}

One of the goals of the proposed algorithm is to have a flexible and lightweight algorithm for adaptive sampling which we demonstrated in this evaluation. Moreover, we target ultra-low power embedded hardware, where the on-time of each component has to be minimized, including the \ac{MCU} itself. We demonstrated that in parity of conditions, the computational overhead is orders of magnitude lower than the \ac{FHC} when run in the same processors.  
A complete iteration from the sampling of the battery voltage using the \ac{ADC} (\(16\times\) oversampling, \qty{40}{\micro\second} acquisition time) until the adjustment of \(k\), the algorithm requires just \qty{706.24}{\micro\second}\footnote{code optimized with \texttt{-O2}} in average, and consuming about \qty{2.70}{\micro\joule} per day. More detailed results can be found in \cref{tab:e_of_algorithm}.\vspace{-0.2cm}

\begin{table}[htpb!]
    \centering
    \caption{Duration and energy consumption over one day of the adaptive sampling algorithm}\label{tab:e_of_algorithm}
    \vspace{-0.2cm}
    \begin{tabular}{@{}lrr@{}}
    \toprule
    \textbf{Task} & \textbf{Duration} & \textbf{E per day}\\
    \midrule
    Sampling the \ac{ADC} & \qty{698}{\micro\second} & \qty{2.57}{\micro\joule}\\
    Estimating charge level & \qty{4.08}{\micro\second}\ (261 cycles) & \qty{61.99}{\nano\joule}\\
    Metric and \ac{FSM} & \qty{4.16}{\micro\second}\ (266 cycles) & \qty{63.18}{\nano\joule}\\
    \midrule
    TOTAL & \qty{706.24}{\micro\second} & \qty{2.70}{\micro\joule}\\
    \bottomrule
    \end{tabular}
    \vspace{-0.5cm}
\end{table}
\section{Conclusion}\label{sec:conclusion}
This paper presented the design and implementation of a low-complexity energy-aware algorithm for adaptive sampling under constrained energy conditions. The algorithm was compared to the state-of-the-art algorithms for adaptive sampling and evaluated on a real-world application use case. Our proposed algorithm has a performance gap from an optimal clairvoyant \ac{FHC} algorithm between $0.95\times$ and $0.85\times$ while being between $2.06\times$ and $8.68\times$ better than a constant sampling rate algorithm tuned on history data. On the other hand, it has an order of magnitude lower computational cost leading to a smaller energy overhead for the algorithm. 
The proposed algorithm has been ported on \textsc{EcoTrack}, demonstrating the self-sustainability of a \ac{GNSS}-enabled \ac{IoT} device over different latitudes, thanks to a solar cell model validated from collected harvesting data.

Future work will involve further evaluation of the effective localization rate when considering the energy overhead in the estimation of the localization rate. We intend to expand the comparison with other adaptive algorithms and incorporate a \ac{LUT} containing various parameters tailored to the geolocation of the device. This approach aims to develop an energy-and-location-aware algorithm, potentially enhancing the performance of the proposed algorithm. 

\begin{acronym}
    \acro{IoT}{Internet of Things}
    \acro{MCU}{microcontroller unit}
    \acro{RF}{radio frequency}
    \acro{SoC}{system-on-chip}
    \acro{BLE}[\textsc{BLE}]{\textsc{Bluetooth Low Energy}}
    \acro{IMU}{inertial measurement unit}
    \acro{GPS}[\textsc{GPS}]{Global Positioning System}
    \acro{GNSS}{global navigation satellite system}
    \acro{UART}{universal asynchronous receiver transmitter}
    \acro{LTE}{Long-Term Evolution}
    \acro{LTE-M}{Long-Term Evolution Machine Type Communication}
    \acro{NB-IoT}{Narrowband \ac{IoT}}
    \acro{A-GPS}{assisted \ac{GPS}}
    \acro{A-GNSS}{assisted \ac{GNSS}}
    \acro{LPWAN}{low-power wide-area network}
    \acro{AWS}{Amazon Web Services}
    \acro{RTOS}{real-time operating system}
    \acro{MQTT}[\textsc{MQTT}]{\textsc{MQTT}}
    \acro{RAM}{random-access memory}
    \acro{RSSI}{received signal strength indicator}
    \acro{RFID}{\ac{RF} identification}
    \acro{SiP}{system-in-package}
    \acro{IC}{integrated circuit}
    \acro{SMU}{source measurement unit}
    \acro{RMSE}{root-mean-square error}
    \acro{MAE}{mean-absolute error}
    \acro{TCP}{Transmission Control Protocol}
    \acro{FHC}{Finite Horizon Control}
    \acro{ADC}{analog-digital converter}
    \acro{FSM}{finite state machine}
    \acro{AIMD}{additive-increase/multiplicative-decrease}
    \acro{MPC}{model predictive control}
    \acro{LUT}{look-up table}
    \acro{WSN}{wireless sensor network}
    \acro{EWMA}{exponentially weighted moving average}
    \acro{LED}{light-emitting diode}
    \acro{DVFS}{dynamic voltage and frequency scaling}
\end{acronym}
\bibliographystyle{ACM-Reference-Format}
\bibliography{cite}
\end{document}